\documentclass[prb,aps,twocolumn,amssymb,amsmath,showpacs]{revtex4}
\usepackage{graphicx}
\newcommand{\figwidth}{.9\columnwidth}

\newcommand{\fig}[1]{Fig.~{\ref{#1}}}
\begin{document}
\title{X-ray spectrum of a pinned charge density wave} \author{Alberto
  Rosso} \author{Thierry Giamarchi} \affiliation{Universit\'e de Gen\`eve,
  DPMC, 24 Quai Ernest Ansermet, CH-1211 Gen\`eve 4, Switzerland}

\begin{abstract}
  We calculate the X-ray diffraction spectrum produced by a pinned
  charge density wave (CDW). The signature of the presence of a CDW
  consists of two satellite peaks, asymmetric as a consequence of
  disorder.  The shape and the intensity of these peaks are determined
  in the case of a collective weak pinning using the variational
  method. We predict divergent asymmetric peaks, revealing the
  presence of a Bragg glass phase.  We deal also with the long range
  Coulomb interactions, concluding that both peak divergence and
  anisotropy are enhanced.  Finally we discuss how to detect
  experimentally the Bragg glass phase in the view of the role played
  by the finite resolution of measurements.
\end{abstract}
\maketitle

The study of disordered elastic objects sheds light on the physics of
a wide range of systems. A first class of systems 
is accurately modeled by elastic manifolds in the presence of randomness; 
significant examples include domain walls in magnetic
\cite{lemerle_domainwall_creep} or ferroelectric
\cite{tybell_ferro_creep} materials, contact lines of liquid menisci
on a rough substrate \cite{rosso_width2} and propagating cracks in
solids \cite{bouchaud_bouchaud}. A second class of disordered elastic
systems is given by periodic structures such as charge density waves (CDWs)
\cite{gruner_revue_cdw}, vortex flux lines in type-II superconductors
\cite{blatter_vortex_review} and Wigner crystals
\cite{chitra_wigner_long}.  It was recently shown that periodic
systems have unique properties, quite different from the ones of the
interfaces. In fact, if topological defects (i.e. dislocations etc.) in the
crystal are excluded, displacements can grow only logarithmically
\cite{nattermann_pinning,korshunov_variational_short,giamarchi_vortex_short},
in contrast with the power-law growth of interfaces. The positional order is
only algebraically destroyed 
\cite{giamarchi_vortex_short,giamarchi_vortex_long}, leading to divergent Bragg
peaks and a nearly perfect crystal state. Quite remarkably, it was shown that
for weak disorder this solution is \emph{stable} to the proliferation of
topological defects, and thus that a thermodynamically stable phase having
both glassy properties and quasi-long range positional order exists
\cite{giamarchi_vortex_long}. This phase, nicknamed {\em Bragg glass}, 
has prompted many further analytical and experimental studies (see e.g.
Ref.~[\onlinecite{nattermann_vortex_review,giamarchi_vortex_review}] for
reviews and further references). Although its existence can be tested
indirectly by the consequences on the phase diagram of vortex flux lines, the
most direct proof is to measure the predicted algebraic decay of the
positional order. Such a measurement can be done by means of diffraction
experiments, using either neutrons or X-rays. Neutron
diffraction experiments have recently provided unambiguous evidences
\cite{klein_brglass_nature} of the existence of the Bragg glass phase for
vortex lattices.

Another periodic system in which one can expect a Bragg glass to occur
are CDWs \cite{gruner_revue_cdw}, where the electronic
density shows a sinusoidal modulation. As a consequence of the 
electron-phonon interaction, this modulations generates a permanent distortion
of the underlying lattice. This distortion can be revealed thanks to X-ray
measurements: in fact, the corresponding X-ray spectrum presents satellite
peaks around each principal Bragg peak. These satellites contain information 
concerning the positional order of the CDW.
In particular, we are interested in the detection of effects due to disorder
\cite{fukuyama_pinning}.
The X-ray experimental resolution is in principle much higher than the one that can be
achieved by neutrons for vortex lattices, consequently CDW systems
should be prime candidates to check for the existence of a Bragg glass
state \cite{rouzieres_structure_cdw}. However, compared to the case of vortex lattices the
interpretation of the spectrum is much more complicated for two main
reasons: (i) the phase of the CDW is the object described by an
elastic energy, whereas the X-rays probe the displacements of the
atoms in the crystal lattice (essentially a cosine of the phase); (ii)
since the impurities substitute for some atoms of the crystal, the very
presence of the impurities changes the X-ray spectrum. This gives rise to
non-trivial terms of interference between disorder and atomic
displacements
\cite{ravy_x-ray_peakasymmetry,rouzieres_structure_cdw}. It is thus
necessary to make a detailed theoretical analysis of the diffraction
due to a pinned CDW. 

In the past, the study of the spectrum has been carried out 
only either for strong pinning or at high temperatures
\cite{rouzieres_structure_cdw,ravy_x-ray_whiteline,ravy_x-ray_peakasymmetry,
brazovskii_x-ray_cdwT}.
In this paper we focus on the low temperature limit, where a well
formed CDW exists, and on weak disorder, for which one expects to be
in the Bragg glass regime. Both the short and long range screening of
the Coulomb interactions are considered.  We show that in both cases
the diffraction spectrum consists in two asymmetric peaks. The peaks
are power-law divergent, with a stronger anisotropic shape in the case
of unscreened long-range Coulomb interaction.  This finding is
consistent with the Bragg glass behavior
\cite{giamarchi_vortex_long}. The asymmetry divergence follows a
subdominant power-law as well, with an exponent that we determine. A
short account of part of the results of this paper was published in
Ref.~\onlinecite{rosso_cdw_short}.

In section~I we derive the model used to describe the interaction
between the CDW and impurities. Two elastic limits are considered: if
free electrons are present the elasticity has a simple short-range
form, while, in the unscreened case, Coulomb interactions are
responsible for a long range strongly anisotropic elastic term.
In section~II we discuss the X-ray intensity spectrum behavior in
presence of a pinned CDW. In particular, we derive the different
contributions to the satellite peaks and we study their symmetry
properties.  In section~III we evaluate explicitly the different terms by means
of the replica techniques.  Section~IV contains the physical
discussion in view of all the results obtained in this paper.  The reader not
interested in the details of calculations may move directly to this
section skipping the previous one.  Finally, in appendix A we
evaluate the triplet contribution and in appendix B we calculate the
function $[\sigma]$, used in section~IV.

\section{The model}

The system we have studied is a  CDW in a three dimensional space.
The electron density has the form \cite{nattermann_brazovskii}:
\begin{equation}
 \rho(r) = \rho_0 +\frac{\rho_1|\psi| Q^{-1}}{\pi}\nabla\phi(r)
 +\rho_1|\psi| \cos( Q r+\phi( r))) \,,
\label{eq:rhoexp}
\end{equation}
where a single sinusoidal deformation of modulation vector
$Q$ is considered and $\psi=|\psi|e^{i\phi}$ is the CDW order parameter normalized
to unity at $T=0$. The first term of Eq.~(\ref{eq:rhoexp}), $\rho_0$, is the average
density. The second one corresponds to a density averaged at scales
larger than $ Q^{-1}$. This contribution, also called {\em
forward scattering}, encompasses the local changes of the electron density related to the
compression modes. The last term, also called {\em backward
scattering}, describes the sinusoidal modulation
at a scale of the order of $Q^{-1}$.  We can neglect all contributions stemming from
higher harmonic terms as they are known to be important only at very low
temperature. The effective Hamiltonian can be obtained by a
Ginsburg-Landau expansion of the order parameter
\begin{equation}
 H=\int d^3 r \frac{1}{2} |\psi|^4 -a |\psi|^2 +\frac{b}{2}|\nabla \psi|^2 \,,
\end{equation}
where $a=(T_c-T)/T_c$ and $b$ is a parameter whose value is defined by the microscopic
theory.  The configuration at minimum energy corresponds to $|\psi|=\sqrt{a}$ with
the phase $\phi$ equal to a constant. Around this equilibrium solution, fluctuations
involve both the amplitude and the phase of the order parameter.  We remark that 
while the first ones are more expensive in terms of energy variations, 
due to the presence of the quadratic term, the second
ones are massless. Following the model developed by Fukuyama-Lee-Rice (FLR)
\cite{fukuyama_pinning,lee_coulomb_cdw,lee_depinning}, we take into account only the
phase fluctuations (i.e. we neglect amplitude fluctuations).  The mass of
these excitations, called phasons, turns out to be quite large, because it 
depends on the ionic mass via the electron-phonon interaction. For this reason,
we can neglect in the Hamiltonian the kinetic term giving rise to quantum fluctuations.
Within these hypotheses, the elastic Hamiltonian associated to the CDW reads
\begin{eqnarray}
\label{eq:elastic}
H_{\text{el}}= \int d^3r \frac{c}{2}  \left( \nabla \phi(r) \right)^2 \,,
\end{eqnarray}
where $c=b |\psi|^2$ is the elastic constant. The thermal fluctuations of the
phase, denoted as $\phi_T$, are easily evaluated:
\begin{eqnarray}
\label{eq:thermal}
{\phi^2_T} &=& 2\langle \phi^2(x)\rangle = \int {\mathcal D}\phi e^{- H/T} \\
&=& 2T\int_{BZ} \frac{1}{cq^2} \sim \frac{Q
T}{\pi^2 c} \,. \nonumber 
\end{eqnarray}
The integral in $q$ extends all over the Brillouin zone.  In the low temperature regime, the
fluctuations are thus small enough to guarantee the presence of an ordered phase.
Moreover, we observe that the value of the integral (\ref{eq:thermal})
is related to the shortest length scale in the problem. Here we
assume that this cut-off momentum, $\Lambda$, is given by the periodicity of the CDW
($\Lambda \sim Q$) \footnote{In the anisotropic case, for a more realistic
  description of the system, we introduce two cut-off wave-vectors. One, along the
  $Q$-direction, remains of the order of $Q$. The others, along the
  transversal directions, are given by the periodicity of the lattice $\Lambda
  = \pi /a$ }.

The form of the Hamiltonian (\ref{eq:elastic}) is actually anisotropic
along the $Q$-direction. In fact, a compression along $Q$ produces an increase
of the electric charge density which yields an increase of the stiffness,
whereas all distortions along the other two directions do not involve any
change in electrostatic energy.

We evaluate the contribution of Coulomb interactions screened beyond the
characteristic length $\lambda$. Without any loss of generality we assume
$Q\parallel x$ and $y$ is along one of the other two equivalent directions. 
The electrostatic
energy takes the form:
\begin{eqnarray}
\label{eq:electr}
U \propto \frac{1}{2 V}  \int d^3r  d^3r'  e^{-|r-r'|/\lambda}
\frac{\rho(r)\rho(r')}{| r- r'|} \,.
\end{eqnarray}
The main contribution to the electrostatic energy comes from the variations
of the electron density $\rho$, expressed by Eq.~(\ref{eq:rhoexp}). 
Therefore, we restrict to consider only the forward scattering term:
\begin{eqnarray}
\label{eq:electr2}
U &\propto& \frac{\rho^2_1 |\psi|^2}{ 2 \pi^2 Q^2 V} \int d^3r  d^3r'
e^{-|r-r'|/\lambda}\frac{\partial_x\phi(r)\partial_{x'}( r')}{| r- r'|}  \nonumber\\
&=& \frac{\rho^2_1 |\psi|^2}{ \pi Q^2 } \int_{BZ}
\frac{(q_x\lambda)^2}{1+\lambda^2 q^2} |\phi(q)|^2
\end{eqnarray}
This term introduces a $q$-dispersion in the elastic constant
 \begin{eqnarray}
 H_{\text{el}}&=& \int_{BZ} \left[c_1(q) q_x^2 + \frac{c}{2} q^2\right]|\phi(q)|^2  \nonumber \\
c_1(q)&\propto& \frac{\lambda^2}{1+q^2\lambda^2} \,. \nonumber 
\end{eqnarray}
Two regimes can be identified as a function of the screening length
$\lambda$. (i) In the first one, valid for $\lambda$ not very large, we
can neglect the dispersion in $q$ and the resulting elasticity is
short-range. The effect of Coulomb interaction is an enhancement of
the elastic constant along the $x$--direction
\begin{eqnarray}
 \label{eq:anisotropy}
 H_{\text{el}}= \int d x d^{2}y \left[ \frac{c_1}{2} \left( \partial_x \phi
 \right)^2 + \frac{c}{2} \left( \partial_y \phi \right)^2 \right] \,.
\end{eqnarray}
By redefining the spatial variables, $x' = x/\sqrt{c_1}$ and $y' = y/\sqrt{c}$,
with $c=(c_1c^{2})^{\frac{1}{2}}$, the Hamiltonian (\ref{eq:anisotropy}) can
be finally turned into an isotropic form. (ii) The second regime is characterized by
large values of $\lambda$. In this case, the electrostatic energy takes the 
form $U \sim \int_{BZ} (q^2_x / q^2) |\phi(q)|^2$, 
and consequently a long-range term appears in the elasticity:
\begin{eqnarray} 
\label{eq:anisotropyfinal}
 H_{\text{el.}}= \int \frac{d q_x}{2 \pi} \frac{d^{2}q}{(2 \pi)^2} \left[ \frac{c_1}{2}
 \frac{q_x^2}{q^{2}} + \frac{c}{2} q^2\right] \left|\phi(q)\right|^2 \,.
\end{eqnarray}

\begin{figure}
\centerline{\includegraphics[width=\columnwidth]{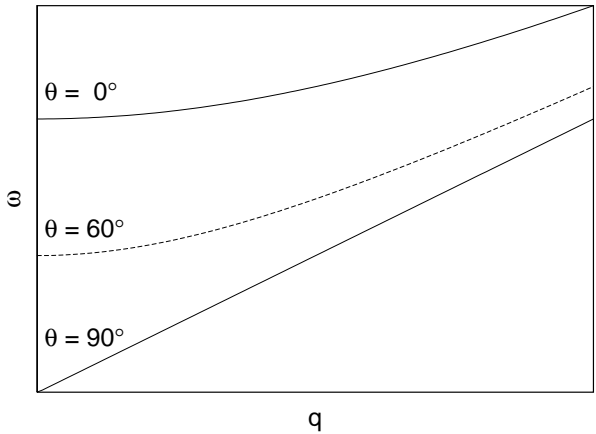}}
\caption{Dispersion relation of the phase mode given by
  (\ref{eq:anisotropyfinal}). The transversal mode ($\theta={90^{\circ}}$) is
  acoustic, the longitudinal mode is gapped ($\theta={0^{\circ}}$). The gap is
  reduced by the presence of longitudinal components ($\theta={60^{\circ}}$).}
\label{omega}
\end{figure}

At this stage, we briefly discuss the dispersion relation of  different elastic regimes.
The full FLR Hamiltonian is:
\begin{eqnarray} 
 H_{\text{FLR}}= \int_{BZ} \frac{1}{2M}P_q P_{-q} +\frac{c(q)}{2} \phi_q \phi_{-q} \,,
\end{eqnarray}
where $P_q$ is the Fourier transform of the momentum density and $M$ is the
phason mass density. The first term gives the
kinetic energy and the second the elastic energy. Using the standard
canonical transformation we derive the corresponding dispersion relation:
\begin{eqnarray} 
 \omega(q)= \sqrt{\frac{c(q)}{M}}\,. 
\end{eqnarray}
If we decompose the vector $q$ in its longitudinal ($q_x= q \cos\theta$) and
transversal ($q_{\perp}=q \cos\theta$) components, it is clear that in the
short-range case $q$ obeys a linear dispersion law, with a slope equal to
$\sqrt{(c_1 \cos(\theta)+c \sin(\theta))/ M}$.  The dispersion for the
long-range elasticity (\ref{eq:anisotropyfinal}) is displayed in \fig{omega}: the
transversal modes remains acoustic, the longitudinal ones instead develop a gap
\cite{gruner_book_cdw}.

Finally, we consider the effect of a distribution of impurities with concentration
$n_I$. The simplest coupling with the electron density is expressed by:
\begin{eqnarray}
\label{eq:disorder}
H_{\text{dis}}= \pm V_0 \int d^3r \Sigma( r) \rho(r) \,
\end{eqnarray}
where $\Sigma(r)$ is the impurity probability distribution. Long range
interactions are neglected and $V_0$ is a positive constant which measures the
impurity potential. At last the sign $+$ ($-$) is related to the repulsive
(attractive) interaction between the electrons and the local impurity.  Above
two dimensions we can drop out the forward
scattering term  in the development of $\rho(r)$.
In fact, this term leads only to a trivial redefinition of the correlation
functions \cite{giamarchi_vortex_long}.  If $V_0$ is small (the opposite case,
the effect of a strong impurity, is discussed in Ref.~[\onlinecite{tutto}])
the FLR model for the elasticity is justified. In this case, the collective pinning
$\Sigma( r)$ is well described by a Gaussian distribution with zero average
(we can always incorporate the effect of the averaged disorder into the bare
parameters) and the correlator is given by $\overline{\Sigma(r)\Sigma(r')} = N_I \delta(r-r')
$,  with $N_I=n_I(1-n_I)$.  
In the following, we will restrict our analysis to the
repulsive case: $\rho_1 |\psi|$ is absorbed in $V_0$ and we define the
disorder strength $D= {V_0}^2 N_I$.

On one hand, the disorder favors local distortions of the
phase $\phi$, on the other hand these deformations 
increase the elastic energy.  A natural size $R_a$
is defined if we consider the region where $\phi$ varies by $2 \pi$. A
simple energetic balance gives for a $d$--dimensional CDW
\begin{equation}
\label{eq:larkin}
E_{tot}=\frac{c}{2} \left(\frac{2 \pi}{R_a}\right)^2 -
\frac{D^{1/2}}{R_a^{d/2}}.
\end{equation}
Optimizing the gain in potential energy versus the cost in elastic
energy we get $R_a= (c^2 / D)^{1/(4-d)}$ (for $d=3$ $R_a =c^2 / D$).
This length, called Fukuyama-Lee length (or Larkin-Ovchinikov length)
\cite{fukuyama_pinning,larkin_ovchinnikov_pinning}, is
interpreted as the correlation length of the system. In this scenario
the equilibrium state is always disordered and the long-range
coherence in the phase is lost.  Nevertheless, in this paper we show that, as in the
case of vortex lattice, the latter prediction is correct only if we consider
scales smaller than $R_a$, while it breaks down at larger distances. In
particular, we see that it is the Fukuyama-Lee length to define the crossover
between the short distance regime and the asymptotic one.

\section{Spectrum Intensity}

The X-ray diffraction is a powerful tool to detect any subtle change of
the perfect crystalline structure.  The electron density modulation is
accompanied, via the electron-phonon interaction, by a lattice distortion $u$
given by
\begin{equation}
\label{displacement}
 u(r) = \frac{u_0}{Q} \nabla \cos(Qr+\phi(r)) \propto \nabla \rho(r).
\end{equation}
Thus, the  CDW
instability produces a permanent sinusoidal displacement of the atoms from
their equilibrium position.  This deformation is signaled,
in the X-ray spectrum, by the presence of satellite peaks around each principal Bragg
peak.  The analysis of the shape, the intensity and the symmetry of such peaks
allows to fully characterize the structural properties of the CDW.  
In this section, we isolate the different terms 
which contribute to the satellite peaks, and study their symmetry properties.  
This discussion is general and model independent.

\begin{figure}
\centerline{\includegraphics[width=\columnwidth]{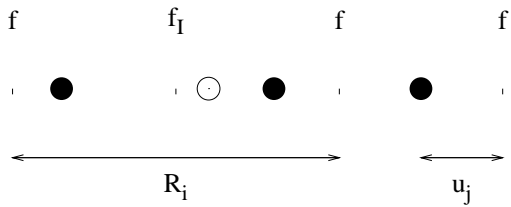}}
\caption{Example of 1--dimensional crystal. The position of the $ith$--atom
  is given by the equilibrium position $R_i$ and the displacement $u_i$.  In
  the figure the black circles are the host atoms with scattering factor $f$
  and the white circle represents an impurity with scattering factor $f_I$}
\label{Itot}
\end{figure}

The expression for the total diffraction intensity of a crystal is:
\cite{guiner_xray}
\begin{eqnarray}
    \label{eq:Sdef}
    I(q) = \frac{1}{V} \sum_{i,j} e^{-iq(R_i-R_j)} \left\langle
\overline{f_if_j e^{-iq(u_i-u_j)}} \right\rangle.
\end{eqnarray}
As sketched in Fig~\ref{Itot}, $u_i$ is the atom displacement from the
equilibrium position $R_i = ia$, with $a$ indicating the lattice constant.
$\overline{\langle\ldots \rangle}$ denotes the double average over the
disorder and over the thermal fluctuations.  $f_i$ represents the total
amplitude scattered by the atom at the position $i$ and depends exclusively on
the atom type.  We consider the simple case of a disordered crystal,
made of one kind of atoms, characterized by the scattering factor $f$, 
and containing impurities of scattering
factor $f_I$. To understand the role of the scattering factors, let us
start by evaluating the case of fixed atoms ($u_i=0$). We obtain:
\begin{eqnarray}
\label{Laue}
I(q) = \overline{f}^2  \sum_K \delta(q-K)
+  {\Delta f}^2 N_I \,,
\end{eqnarray}
where $\Delta f= f_I -f $ and $\overline{f}$ is the average
scattering factor.  The usual Bragg peaks, corresponding to the reciprocal
lattice vectors $K$, arise from the first term in (\ref{Laue}), while the second
term is responsible for a constant background intensity, called {\em Laue
scattering}, due to the disorder.

Now we move back to the general case $u_i \ne 0$. 
We are interested in the behavior of the scattering intensity $I(q)$ near a
Bragg peak ($q \sim K$). Since $|\delta q| =|q-K| \ll K$, we can take the
continuum limit $i \rightarrow r$ and obtain from (\ref{eq:Sdef}):
\begin{equation}
    \label{eq:Sdef2}
    I(q) =  \int_{r_1,r_2}  e^{-i \delta q
  (r_1-r_2)} \left\langle
\overline{f_{r_1}f_{r_2}
 e^{-i \delta q (u(r_1)-u(r_2))}} \right\rangle \,.
\end{equation}
Here $\int_{r_1,r_2} = \frac{1}{V a^{d}} \int d^dr_1 d^dr_2$ and $f_r =
\overline{f}+ \Delta f a^{d/2} \Sigma(r)$.  Assuming that in the elastic
approximation the displacements remain small ($u_i \ll R_i$), one can expand
(\ref{eq:Sdef2}) as a power series of $K u_0$.  Developing up to the second
order we get \cite{ravy_x-ray_whiteline} :
\begin{widetext}
\begin{equation}
\label{eq:Intensities}
  I(q) = I_{\text{d}} + I_{\text{a}}+ I_{\text{tripl}}, \mbox{with} 
\end{equation}
\begin{eqnarray}
 I_{\text{d}} =\overline{f}^2 q^2  \int_{r_1,r_2}  e^{-i \delta q (u(r_1)-u(r_2))}
 \langle \overline{u(r_1) u(r_2)} \rangle, &&
I_{\text{a}} =  -iq {\Delta f} \int_{r_1,r_2}
  e^{-i \delta q (u(r_1)-u(r_2))}  \left\langle
\overline{\Sigma(r_1) u(r_2) - \Sigma(r_1)
 u(r_2) } \right \rangle, \nonumber 
\end{eqnarray}
\begin{equation}
 I_{\text{tripl}}= - iq {\Delta f}^2 a^{d} \int_{r_1,r_2}  e^{-i \delta q
     (u(r_1)-u(r_2))} \left\langle \overline{\Sigma(r_1)\Sigma(r_2) (u(r_1)
- u(r_2)) } \right \rangle \,. \nonumber
\end{equation}
\end{widetext}
While the contribution $I_{\text{d}}$ represents the intensity due to the
atomic displacements alone, the contributions $I_{\text{a}}$ and
$I_{\text{tripl}}$ are generated by the coupling between the disorder and the
displacements.  In the following we consider only $I_{\text{a}}$ and
$I_{\text{d}}$; the term $I_{\text{tripl}}$ is evaluated in the appendix
\ref{appendice} where we show that it is smaller than the other two.

In a pure system ($\Sigma(r)=0$) we expect that only the first term can be
different from zero.  Referring to the case of a CDW without disorder and
neglecting the thermal fluctuations, we can replace in (\ref{eq:Intensities}) the
displacement term (\ref{displacement}) with $\phi(r)=\text{const}$.  This gives:
\begin{equation}
 I_{\text{d}}(q) = f^2 q^2 \sum_K \delta(q \pm Q -K).
\end{equation}
The presence of two satellites around each Bragg peak is one of the most clear
experimental evidence of a CDW. In a pure system these satellites are
symmetric and without any broadening.  To interpret the experimental findings
\cite{ravy_x-ray_peakasymmetry,brazovskii_x-ray_cdwT,rouzieres_structure_cdw},
which reveal the presence of asymmetric peaks,
we need to account for the effect of impurities. In particular, we want to
explain not only the measured intensity asymmetry (IA) between the two satellites, but
also the profile asymmetry (PA) of each peak, which is measured in
strongly doped samples.

The symmetry properties of the terms $I_{\text{a}}$ and $I_{\text{d}}$ can be
determined by considering the lattice displacements and the disorder expressed
in terms of their Fourier components: 
\begin{eqnarray}
\label{FT}
 u(r) & = & \int_{BZ} e^{-i q r} u_q \nonumber  \\
\Sigma(r)&=& \int_{BZ} e^{-i q r} \Sigma_q \,.
\end{eqnarray}
It easy to obtain:
\begin{eqnarray}
\label{Ftermsd}
 I_{\text{d}}(q) & = & \overline{f}^2 q^2  \overline{\langle u_{\delta q}
 u_{-\delta q}  \rangle}\\
\label{Ftermsa}
I_{\text{a}}(q) & = & 2  a^{d/2} q \Delta f \overline{f} \overline{\langle
Im[\Sigma_{-\delta q} u_{\delta q}] \rangle}.
\end{eqnarray}
\begin{figure}[t]
\includegraphics[width=\columnwidth]{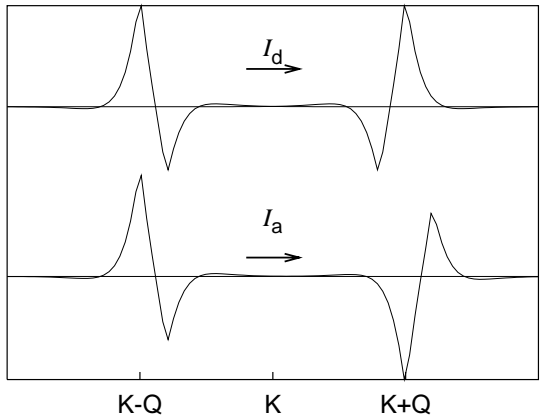}
\caption{Sketch showing the symmetry properties of the two different terms
  contributing to the satellite intensities (the Bragg peak at $K$ is not
  shown).  Top: term $I_{\text{d}}$. The intensity is symmetric with respect
  to the Bragg peak.  Bottom: term $I_{\text{a}}$. The intensity is
  antisymmetric respect to the Bragg peak.}
\label{Isym}
\end{figure}

The two prefactors, $\overline{f}^2 q^2$ and $\Delta f \overline{f} q$ vary
slowly in $q$ and one can assume they are constant in the neighborhood of the reciprocal
lattice vector $K$. Due to the fact that (\ref{Ftermsa}) is an imaginary part, we can deduce:
\begin{eqnarray}
\label{symmetry}
 I_{\text{d}}(K+\delta q) & = &  I_{\text{d}}(K-\delta q) \nonumber \\
 I_{\text{a}}(K+\delta q) & = & - I_{\text{a}}(K-\delta q) \,.
\end{eqnarray}
In \fig{Isym} we sketch the following symmetry properties of different terms:
$I_{\text{d}}$ generates two satellites symmetric with respect to the Bragg peak,
while $I_{\text{a}}$ gives antisymmetric contributions.  We conclude that the  IA is due to
$I_{\text{a}}$ while the PA is not excluded in both terms, in particular we
have a mirror symmetry for $I_{\text{d}}$. Plugging  the
displacement form (\ref{displacement})  given by the FLR model
in (\ref{Ftermsd}), we get
\begin{equation}
 \label{Id}
 I_{\text{d}}(K + Q + k)  =  u_0^2 f^2 K^2 \int_r e^{-i k r}
 C_{\text{d}}(r) \,,
\end{equation}
where $\int_r = \frac{1}{ a^{d}} \int d^dr $ and $C_{\text{d}}(r)$ is the
positional correlation function
\begin{equation}
 \label{Cd}
 C_{\text{d}}(r)  = \left\langle \overline
 {e^{i(\phi(\frac{r}{2})-\phi(-\frac{r}{2}))}} \right \rangle \,.      
\end{equation}
Examining this equation one is able to connect the symmetry properties of the
satellite peak profile with the symmetry properties of the system.  In
particular, if $C_{\text{d}}(r)$ is an even function in $r$, the $I_{d}$ term
cannot show a PA. On the other hand $I_{d}(q)$ is a real function so the
latter condition is equivalent to require the presence of a $\phi \to -\phi$ symmetry in the
system. We will make use these observations to interpret the experimental findings in
section \ref{physics}.

\section{The replica method}

In this section we calculate the different terms of the development of
Eq.~(\ref{eq:Intensities}) using a Gaussian variational approach
\cite{mezard_variational_replica,giamarchi_vortex_long}.  We consider first the
isotropic case, to include later the corrections due to the Coulombian
interactions. The physical interpretation of these results is presented in more detail in
section \ref{physics}.

We consider a FLR Hamiltonian $H=H_{\text{el}}+H_{\text{dis}}$ where the
second term is (\ref{eq:disorder}) and the first one is described either by
(\ref{eq:elastic}), in case of a short range elasticity, or by (\ref{eq:anisotropyfinal})
taking into account the effect of long-range Coulombian interactions.
We first perform the average over the disorder using the replica techniques.
The replicated Hamiltonian is:
\begin{equation} \label{eq:ham}
 H_{\text{\text{eff}}}= \sum_a H^a_{\text{el}} -\int d^dr\frac{D}{T} \sum_{a,b}
 \cos\left(\phi_a(r) -\phi_b(r)\right) \,,
\end{equation}
where $T$ is the temperature and the sum over the $n$ replicas has to be
considered in the limit $n \rightarrow 0$. We observe that the system is $\phi
\to -\phi$ invariant. This means that the FLR model with a Gaussian disorder
cannot generate satellite peaks with a PA. Finally, we stress that, moving from
the original Hamiltonian to its replicated version we also need to change the
correlation functions containing explicitly the disorder.  In particular,
using (\ref{displacement}), the terms $I_{\text{d}}$ and $I_{\text{a}}$ obtained
in (\ref{eq:Intensities}) become
\begin{eqnarray}
 I_{\text{d}} & = & F_{\text{d}}\int_{r_1,r_2}e^{-i \delta q
  (r_1-r_2)}
 \nabla_{r_1}\nabla_{ r_2} 
 \langle  \rho_1(r_1)  \rho_1(r_2)  \rangle_{\text{eff}}  \nonumber \\
 I_{\text{a}}& =& F_{\text{a}}
   \int_{r_1,r_2} e^{-i \delta q
  (r_1-r_2)}  \frac{\nabla_{r_1}-\nabla_{ r_2}}{n}\sum_{a,b}^n \left\langle
 \rho_a(r_1) \rho_b(r_2) \right \rangle_{\text{eff}} \,,  \nonumber
\end{eqnarray} 
where $F_{\text{d}}= \frac{\overline{f}^2 q^2 u_0^2}{Q^2}$ and
$F_{\text{a}}= \frac{iq u_0 D a^{d/2} \Delta f \overline{f}}{Q T V_0
}$.  Replacing the backward scattering term of the electron density
and performing an integration by parts we get to the form:
\begin{eqnarray} \label{eq:Correlations}
 I_{\text{d}}  &=&  \overline{f}^2 q^2 {u^2}_0 \int_r e^{-i \delta  q r}\left[ e^{-i  Q
 r} +c.c.\right] C_{\text{d}}(r) \,,
   \\
 I_{\text{a}} &=& - \overline{f}  {\Delta f} q  u_0 \sqrt{N_I  a^{d} D }
\int_r e^{-i \delta  q r} \left[ e^{-i Q r} - c.c.\right]
 C_{\text{a}}(r)  \,. \nonumber
\end{eqnarray}
\begin{eqnarray}
C_{\text{d}}(r)&=&\frac{1}{n} \sum_{a} \left\langle e^{i(\phi_a(\frac{r}{2})
    -\phi_a(-\frac{r}{2}))}\right\rangle_{\text{eff}} \nonumber \\
C_{\text{a}}(r)
&=&\frac{1}{Tn} \sum_{a,b}^n \left\langle e^{i(\phi_a(\frac{r}{2})
    -\phi_b(-\frac{r}{2}))} \right \rangle_{\text{eff}}
 \nonumber 
\end{eqnarray}
are the positional correlation functions controlling the behavior of each
contribution.  To obtain this result we have applied the standard decomposition in
center of mass $R$ and relative $r$ coordinates ($r_1 = R+\frac{r}{2}$ and
$r_2 = R-\frac{r}{2}$). Since $u$ varies slowly at the scale of the lattice
spacing, we performed the integration over $R$. We notice that
(\ref{eq:Correlations}) reveals clearly the presence of two peaks situated at $q=Q+K$
and $q=K-Q$.  In particular, as expected, the contribution to the two
satellites of the displacement term $I_{\text{d}}$ is symmetric, while the one of
$I_{\text{a}}$ is antisymmetric. The sum of these two terms leads to the IA
experimentally observed.

Following the same method used to study the flux lines in presence of a weak
disorder \cite{giamarchi_vortex_long}, we can evaluate the different terms in
(\ref{eq:Correlations}). We look for the best trial Gaussian Hamiltonian $H_0=
\int_q G^{-1}_{ab}(q) \phi_a(q) \phi_b(-q)$ in the replica space which
approximates (\ref{eq:ham}).  The $G^{-1}_{ab}(q)$ is the $n \times n$
variational matrix. Without loss of generality, this matrix can be chosen of
the form $G^{-1}_{ab}=c q^2 \delta_{a b}-\sigma_{a b}$. The connected part is
defined as $G^{-1}_{c}=\sum_bG^{-1}_{a b}$. By minimization of the variational
free energy we derive that $G_c$ is given by the bare elastic
propagator. In the isotropic case we write
\begin{equation}
 G_c^{-1} = c q^2 \,.
\label{eq:gcsr}
\end{equation}
For a long range elasticity, in $d$-dimension, it follows:
\begin{equation}
 G_c^{-1} = c_1 \frac{q_x^2}{q^{d-1}} + c q^2 \,.
\label{eq:gclr}
\end{equation}
Finally, the parameters $\sigma_{ab}$ are given by:
\begin{eqnarray}
\label{eq:replica1}
 \sigma_{a \ne b}=\frac{D}{T} e^{-\frac{B_{a b}(r=0)}{2}} \,,
\end{eqnarray}
where
\begin{eqnarray}
\label{eq:replica2}
 B_{ab}(r)&=& \langle (\phi_a(r) -\phi_b(0))^2 \rangle_0 \nonumber \\
          &=&2 T\int_{BZ} \left[\tilde{G}(q) -G_{ab}(q)
 \cos qr \right] \,. 
\end{eqnarray}

$\tilde{G}$ is the diagonal element of $G_{ab}$. In the Gaussian
approximation, the positional correlation functions become:
\begin{eqnarray}
\label{eq:Var-Corr1}
 C_{\text{d}}(r)&=&  e^{-  \frac{\tilde{B}(r)}{2}}  \\
\label{eq:Var-Corr2}
 C_{\text{a}}(r) &=&\frac{1}{Tn} \sum_{a,b}^n e^{-  \frac{B_{a b}(r)}{2}} \,,
 \end{eqnarray} 
where $\tilde{B}$ is the diagonal element of $B_{ab}$. In order to simplify
the notation we write (\ref{eq:Var-Corr2}) as
$C_{\text{a}}(r)=\chi(r)C_{\text{d}}(r)$. Finally, using (\ref{eq:replica2}), we get:
\begin{eqnarray}
\label{RSB}
\chi(r) = \frac{1}{nT}\sum_{a,b}^n e^{-T \int_q (\tilde{G}(q) -
    G_{ab}(q)) \cos qr} \,.
\end{eqnarray}

Two general classes of solutions exist for this problem: while the first class preserves the
permutation symmetry of the replicas (RS), the second class (RSB) breaks the
replica symmetry. It has been shown \cite{giamarchi_vortex_long} that the
stable solution for $d>2$ corresponds to the RSB class, while the RS solution
remains valid at short distance. In the following we will refer directly to the
$d=3$ case.

\subsection{Replica symmetric solution}

We discuss first the RS solution which gives the correct evaluation of the
correlation functions at a distance smaller then $R_a$. Within this Ansatz the
matrix $G$ and $B$ are defined by two of their elements: the diagonal values
$\tilde{G}$, $\tilde{B}$ and the off-diagonal ones, $B = B_{a \ne b}$ and
$G=G_{a \ne b}$. The simple algebra of symmetric matrix yields, for $n=0$,
\begin{eqnarray}
\label{eq:RSGd}
 \tilde{G}&=&G_c(1+G_c \sigma_{a \ne b}) \,,  \\
\label{eq:RSGnd}
 \tilde{G} -G&=& G_c \,,
\end{eqnarray} 
where from (\ref{eq:replica1}) and (\ref{eq:thermal}), we deduce that  $\sigma_{a \ne
  b}\sim D/T \exp(-\phi^2_T/2)$.

We focus first on the short range elasticity. Replacing the form of $G_c$ given in
(\ref{eq:gcsr}) we easily calculate the displacement
\begin{eqnarray}
\label{eq:rsshort}
\tilde{B}(r)  &\sim& \phi^2_T + \frac{4r}{3 \pi^2 R_a}.
 \end{eqnarray}
The first term take into account the thermal fluctuations and saturates to
$\phi^2_T$ at a large distance.  The second term, due to the disorder, grows with
a power law and it is responsible for the exponential decay of the positional
correlation functions. To characterize the spectrum it remains to calculate
\begin{eqnarray}
\label{eq:chirs}
 \chi(r)&=&\frac{1}{T} \left[1 - e^{ \frac{ - 2 \pi^2 T}{ c
             r} } \right]\nonumber \\ &\sim& \frac{2 \pi^2}{ c r}.
 \end{eqnarray}
We can conclude that in the RS scenario the two positional correlation
functions decay exponentially fast, moreover $C_{\text{a}}$ has a power law extra
factor.

Before calculating the stable RSB solution, we evaluate the scaling
behavior of the same objects in the case of an unscreened Coulombian
potential. The only change consists in taking the connected propagator
given in (\ref{eq:gclr}) to determine the physical quantities.
It is instructive to
discuss first the general $d$--dimensional case. The displacement takes the
form 
\begin{eqnarray} \label{eq:displ2}
\tilde B(r)=\int d^{d-1}q dq_x \frac{q^{2d-2}} {\left[c_1 q_x^2 + c
q^{d+1}\right]^2} [1-\cos(qr)] \,.
\end{eqnarray}
A similar integral was discussed in Ref.~[\onlinecite{ledoussal_mglass_long}].
A general remark is that $q_x$ scales as $q^{(d+1)/2}$. The scaling behavior
of these integrals is determined by small $q$'s, for this reason we can neglect
the $q_x$ dependence in $q$. The strong anisotropy  along $x$ and $y$
 can be studied performing the following substitution:
\begin{eqnarray} \label{eq:substitution}
v=xq_x \;\;\; {,} \;\;\; t=qy \nonumber\\
z=\sqrt{\frac{c}{c_1}} \frac{x}{y^{\frac{d+1}{2}}} \,.
\end{eqnarray}
We obtain
\begin{eqnarray} \label{eq:displd}
\tilde B(r) &=& \frac{y^{\frac{9-3d}{2}}}{\sqrt{c^3c_1}}\int d^{d-1}t
 \frac{dv}{z} \frac{t^{2d-2}}{\left[ (\frac{v}{z})^2 +
 t^{d+1}\right]^2} [1- \cos(t+v)] \nonumber\\
 &=&\frac{y^{\frac{9-3d}{2}}}{\sqrt{c^3c_1}} H_1(z) \,,
\end{eqnarray}
where $H_1(0)=const.$ and $H_1(z\rightarrow \infty) \propto z^{(9-3d)/(d+1)}$.
It easy to check that, for $d=3$, there are only logarithmic divergences:
\begin{eqnarray} \label{eq:displfinal}
\tilde B(x=0,y) &\sim& \frac{D}{(2\pi)^2\sqrt{c^3c_1}} \log(\Lambda y) \,,
 \nonumber\\ 
 \tilde B(x,y=0) &\sim& \frac{D}{2(2\pi)^2\sqrt{c^3c_1}} \log(\Lambda_x x) \,.
 \nonumber 
\end{eqnarray}
This result is a clear evidence that, because of the long range interactions,
the system is more rigid and the critical upper dimension becomes $d=3$, in
contrast with the result $d=4$ for the short range case. To confirm this
statement we evaluate the Fukuyama-Lee length by imposing $\tilde
B(r)=(2\pi)^2$:
\begin{eqnarray} \label{larkin}
\tilde R_a(x=0,y) &\sim& \Lambda^{-1} e^{\frac{(2\pi)^4\sqrt{c^3c_1}}{D}} \,,
 \nonumber\\ 
 \tilde R_a(x,y=0) &\sim& \Lambda_x^{-1} e^{\frac{2(2\pi)^4\sqrt{c^3c_1}}{D}} \,.
\end{eqnarray}
The exponential law is characteristic of the upper critical dimension and is
an extrapolation of the power law. The equivalent form (\ref{eq:crossuc}) has
been derived for an isotropic short-range elasticity in $d=4$
\cite{chitra_vortex_upper,bucheli_frg_secondorder}.  It remains to determine
Eq.~(\ref{eq:chirs}). At low temperature we write
\begin{eqnarray}\label{chiani}
\chi(r) &\sim& \int d^dq q^{d-1}\frac{\cos qr}{c_1 q_x^2 + c q^{d+1}}  \nonumber\\
 &=& \frac{y^{\frac{5-3d}{2}}}{\sqrt{cc_1}}\int d^{d-1}t
 \frac{dv}{z} \frac{t^{d-1}}{ (\frac{v}{z})^2 + t^{d+1}} \cos(t+v) \nonumber \\
&=& \frac{y^{\frac{5-3d}{2}}}{\sqrt{cc_1}} H_2(z) \,,
\end{eqnarray}
where $H_2(0)=const.$ and $H_2(z\rightarrow \infty) \propto
z^{(5-3d)/(d+1)}$. 
In $d=3$ a straightforward calculation confirms this scaling
behavior:
\begin{eqnarray} 
 \chi(x,y=0)  &\sim&   \frac{1}{16 \pi c  x}  \nonumber\\
 \chi(x=0,y)  &\sim&   \frac{1.1}{2 \pi^2 \sqrt{c c_1} y^2} \,. \nonumber 
\end{eqnarray}
We observe that the anisotropic scaling $x\sim y^2$ is always verified.  Since
for $d=3$ the RS solution is unstable, to obtain the physics at large distance
one has to consider the RSB method.

\subsection{Replica symmetric breaking solution}

Within this scheme, the off diagonal elements of $G_{ab}(q)$ are
parameterized by $G(q,v)$ where $0<v<1$.  The saddle point equation
becomes
\begin{eqnarray}
\label{eq:replica1b}
 \sigma(v)=\frac{D}{T}  e^{-\frac{B(r=0,v)}{2}} \,.
\end{eqnarray}
We look for a solution such that $\sigma(v)$ is constant above a variational
breakpoint $v_c$.  This can be done in a easy manner by
recasting the equations in terms of a new variable
\begin{eqnarray}
\label{eq:sigmaapp}
[\sigma](v)=v\sigma(v)-\int du \sigma (u) \,.
\end{eqnarray}
It is not difficult to show that $[\sigma]'(v)=v\sigma'(v)$.  We refer to appendix
\ref{appendix:sigma}, where we summarize the previous results for $[\sigma]$ and we
calculate its form in the case of an unscreened Coulombian elasticity. As a
first step one uses the inversion rules of hierarchical matrices
\cite{mezard_variational_replica} , Eqs.~(\ref{eq:RSGd}) and (\ref{eq:RSGnd})
become
\begin{eqnarray}
\label{eq:RSBGd}
 \tilde{G}&=&G_c(1+\int \frac{d v}{v^2}\frac{[\sigma]}{G^{-1}_c+[\sigma]})  \\
\label{eq:RSBGnd}
 \tilde{G} -G&=& \left[ \frac{1}{G^{-1}_c+\Sigma} +
 \int_{v}^{v_c} dt
 \frac{\sigma'(t)}{\left(G^{-1}_c+[\sigma](t)\right)^2}\right] \,, 
\end{eqnarray} 
where $\Sigma=[\sigma](v_c)$ is a variational parameter, whose expression is determined in 
appendix \ref{appendix:sigma}.

Starting from the short range case we calculate the displacement using 
Eq.~(\ref{eq:RSBGd}), replacing the expression (\ref{eq:sigmares}) for $[\sigma]$.
\cite{giamarchi_vortex_long}:
\begin{eqnarray} \label{displRSB}
\tilde{B}(r)  &\sim& \phi^2_T + 4 \pi^2 \int_{BZ} \frac{[1-\cos qr]}{q^3}
 \nonumber\\ 
 &\sim& \phi^2_T + 2 \log(\Lambda r) \,.
\end{eqnarray}
The logarithmic behavior
\cite{giamarchi_vortex_short,korshunov_variational_short} of (\ref{displRSB})
is controlled by small $v$ ($v<v_c$). Values of $v$ above the breaking point
($v>v_c$) give the small distance contribution.  To fully characterize the
spectrum it still remains to evaluate $\chi(r)$ in the RSB scenario.
\begin{equation}\label{RSB3d}
 \tilde{G}(q)-G(q,v) \sim \frac{2}{c l^2}\int_{v/v_c}^1 dt
 \frac{1}{\left(q^2 +(t/l)^2
 \right)^2} \,,
\end{equation}
where the parameters $l$ and $v_c$ are given by (\ref{eq:cross}). 
By integrating (\ref{RSB3d}) over $q$ and with some manipulations,
we get:
\begin{eqnarray}
\label{eq:chirsb}
 \chi(r) &=& \frac{v_c}{T} \left[ 1- \int_0^{1} dz \exp \left(-8\pi^3
 \int^{1}_z \frac{dt}{t} e^{-rt/l} \right) \right] \,.
\end{eqnarray}
\begin{figure}
\centerline{\includegraphics[width=\figwidth]{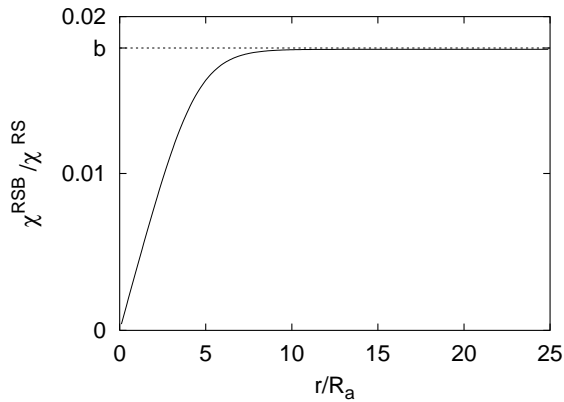} }
\caption{Ratio between the RSB and RS solutions of $\chi(r)$. At
  large distance this ratio tends towards a constant value $b$, with
  $b \sim 0.018$.  This means that the RSB solution affects $\chi(r)$
  only by a multiplicative factor.}
\label{integral}
\end{figure}
The low temperature behavior ($l \sim R_a$) of this term is sketched
in Fig.~\ref{integral}.

Finally we study the RSB solution for the Coulombian elasticity.  Replacing
 (\ref{eq:sigmalrfinal}) in (\ref{eq:RSBGd}), the diagonal correlator becomes
\begin{eqnarray}
\label{eq:tildeGlast}
\tilde{G}&\sim&  G_c\int \frac{d v}{v}\frac{1}{G^{-1}_c \log^2(\frac{A_{lr} v
        }{ \sqrt{c_1 c} 
        \Lambda_x})+ A_{lr} v}  \nonumber \\
&\sim& G^2_c \log^{-1}(\frac{G^{-1}_c}{ \sqrt{c_1 c} \Lambda_x}) \,.
\end{eqnarray} 
The latter form is characteristic of the upper critical
dimension. Inserting Eq.~(\ref{eq:tildeGlast}) in
Eq.~(\ref{eq:replica2}), taking $c_1=c=1$, and employing the usual
substitution (\ref{eq:substitution}) we get:
\begin{equation} 
\tilde B(r) = S_2\int dt
 \frac{dv}{z} \frac{t^{5}}{\left[ (\frac{v}{z})^2 +
 t^{4}\right]^2} [1- \cos(t+v)] \,. \nonumber
\end{equation}
This equation  leads to the same conclusions discussed for the isotropic
upper critical dimension. The asymptotic displacement is thus given by
\begin{eqnarray} \label{displlongrangeRSB}
\tilde{B}(x)  &\sim& \log(\log(\Lambda_x x)) \,.
\end{eqnarray}
As in the case of the short-range elasticity, it can be shown  that the RSB
solution does not affect the asymptotic power low  behavior of $\chi(r)$.

\section{\label{physics} Physical discussion}

In this section we summarize the results obtained in the previous
sections and compare them with the experimental findings.

\begin{figure}
\centerline{ \includegraphics[width=\figwidth]{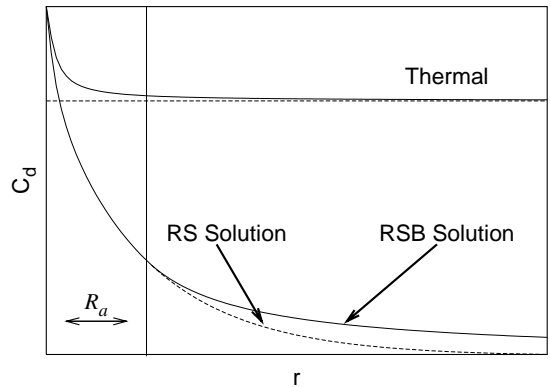} }
\caption{Behavior of $C_d(r)$. The thermal contribution saturates to
$\exp(-\phi^2_T/2)$, the RSB solution is valid beyond $R_a$, while the
RS one is valid below $R_a$.}
\label{C}
\end{figure}
We start by evaluating for the short-range elasticity model the positional
correlation function $C_{\text{d}}(r)$, defined in Eq.~(\ref{Cd}). This
function is the analog of the correlation function determined for vortex line
systems \cite{giamarchi_vortex_long}.  It is well known that for $d>2$ it
exists a finite temperature (the critical temperature $T_c$) below which
thermal fluctuations are prevented from disordering the system. In fact, the
direct calculation of the positional correlation function in $d=3$ yields:
\begin{equation}
C^{therm.}_{\text{d}}(r)= e^{\frac{\phi^{2}_T}{2} (1 -\frac{{\text Si}(\Lambda r)}{\Lambda r})} \,,
\end{equation}
with ${\text Si}(\Lambda r)= \int_0^{\Lambda r} dt \sin t /t $. As shown in
\fig{C}, the correlation function saturates after a few lattice parameters to
a non-zero value, which witnesses the presence of long-range order in the
system.

However, the quenched noise originated in the impurities is still able to
destroys the long range order, even in $d=3$ (see \fig{C}). In this
circumstances, the more traditional scheme \cite{efetov_larkin_replicas} 
describes the system as organized in ordered domains 
(called `Fukuyama-Lee-Rice domains'), characterized by the average size $R_a$.
Beyond this characteristic length, the CDW dislocations become dominant and
any order disappears. This scenario is captured by the RS solution.
From (\ref{eq:rsshort}) we obtain:
\begin{equation}
\label{shortrangesolution}
C^{RS}_{\text{d}}(r)  =   C^{therm.}_{\text{d}}(r) e^{-\frac{4 r}{3 \pi^2
    Ra}} \,.  
\end{equation}
As a direct consequence, we expect to find Lorentzian satellites, whose
half-width at half-height is of the order of $R^{-1}_a$.  However, the RS
approach is unstable at large distances and the correct solution is given
asymptotically by the RSB. Within this assumption, the correlations decrease
following a power law:
\begin{equation}
C^{RSB}_{\text{d}}(r)  =    C^{therm.}_{\text{d}}(r)  (\frac{R_a}{r})^{\eta} \,. 
\end{equation}
Applying the variational approach \cite{eta_exponent_variation}, one
finds, from (\ref{displRSB}), $\eta=1$. The corresponding
quasi-ordered phase
\cite{giamarchi_vortex_short,giamarchi_vortex_long}, called Bragg
glass, is characterized by an infinite correlation length and the
characteristic size $R_a$ represents now the crossover between the RS
and RSB solutions.  One can give a simple physical interpretation of
the two identified regimes, by observing that at the scale $R_a$ the
phase distortions are of the order of the CDW period $2 \pi$.  This
means that for distances smaller than $R_a$ the development of the
Hamiltonian (\ref{eq:ham}) is allowed and leads to the RS
solution. For distances larger than $R_a$, instead, the phase feels
its periodic nature and this trivial development of the Hamiltonian is
no more valid.

\begin{figure}
\centerline{ \includegraphics[width=\figwidth]{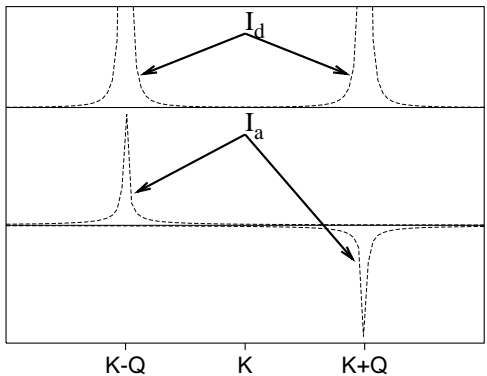} }
\caption{Intensities of the different contributions to satellite
  peaks. The more divergent term, $I_{\text{d}}$, is symmetric.
  $I_{\text{a}}$ is antisymmetric. In this figure we consider a
  repulsive potential and $\Delta f>0$.}
\label{modello}
\end{figure}

In order to determine the intensity and the shape of the two satellites we
need to evaluate all the terms contributing to the development
(\ref{eq:Intensities}). In particular, we consider the displacement term
$I_{\text{d}}$, and the asymmetric term $I_{\text{a}}$, arising from the
coupling between disorder and displacement.  In the literature this latter
term was previously estimated by means of models
\cite{ravy_x-ray_whiteline,ravy_x-ray_peakasymmetry,brazovskii_x-ray_cdwT}
which describe the pinning imposing a constant value $\phi_0$ on the phase in
proximity of each impurity. According to these approaches, the observed
satellite asymmetry is a clear signature of strong disorder
\cite{ravy_x-ray_peakasymmetry}.  Thanks to our more accurate calculation, we
found that the term $I_{\text{a}}$ is non-zero also in case of weak
disorder and it gives rise to a divergent contribution similar to the
one stemming from $I_{\text{d}}$.  Using (\ref{displRSB}),
(\ref{eq:chirs}) and (\ref{eq:chirsb}), our final results read:
\begin{eqnarray}
\label{stima2}
I_{\text{d}}(K+Q+k) & = & \overline{f}^2 K^2 {u_0}^2 \int_r e^{-i k r}
 (\frac{R_a}{r})^{\eta} \,,\\ I_{\text{a}}(K+Q+k) & = & - 2\pi^2
 \overline{f} {\Delta f}  \sqrt{\frac{N_I a}{R_a }} K u_0 \int_r e^{-i k r}
 (\frac{R_a}{r})^{\eta} \frac{ba}{r}\nonumber \,.
\end{eqnarray}

After computing the $d$-dimensional Fourier Transforms, we conclude that both
terms are divergent: in particular, $I_{\text{d}} \propto
\frac{1}{q^{d-\eta}}$ and $I_{\text{a}} \propto \frac{1}{q^{d-\eta-1}}$.  This
result, summarized in Fig.~\ref{modello}, is a clear sign of the
quasi-long range positional ordered phase. In this particular case,
the peak at $K+Q$ is smaller than the $K-Q$ one, since the specific
interaction between the impurity and the CDW is repulsive (we would
have the opposite asymmetry in case of an attractive interaction). We
observe that for an ideal experiment with infinite resolution the
symmetric term would be dominant, as $C_{\text{d}} (r)$ decays to zero
less rapidly than $C_{\text{a}}(r)$.  However, in a real measurement
the divergence in (\ref{stima2}) is always cut by the finite
resolution and both terms have to be accounted for, as the prefactor
$K u_0$ of the less divergent term $I_{\text a}$ is larger.

\begin{figure}
\centerline{ \includegraphics[width=\figwidth]{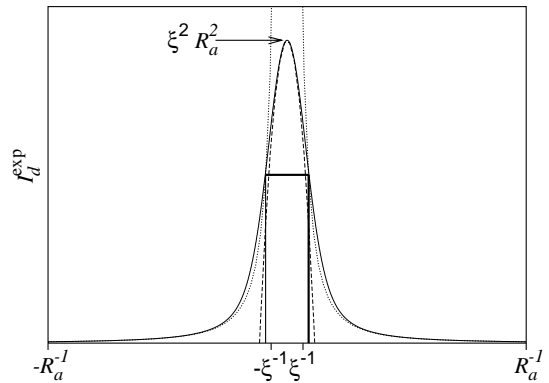} }
\caption{
 Dotted line: divergent peak $I_d(q) =q^{-2}$ for a perfect
 experimental resolution. Full line: peak convolved with a finite
 experimental resolution according to (\ref{last}). The height of the
 peak is $\sim \xi^2 R_a^2$, the half-width half-height $\sim
 \xi^{-1}$ is also indicated. Dashed line: an approximate form for the
 peak $I^{\text approx}_{d}(q) = I_d^{\text exp}d(0)(1- q^2 \xi^2
 /3)$.}
\label{ultima}
\end{figure}

At this stage, following the analysis of the experimental data of
Ref.~[\onlinecite{klein_brglass_nature}] concerning neutron
diffraction spectra, it is interesting to discuss the role played by
the experimental resolution in determining the peak shape.  With the
low resolution achieved by means of neutrons to describe a vortex
lattice, in the above cited experiment \cite{klein_brglass_nature}, it
is possible to determinate the intensity spectrum only along one
direction, after performing an integration on the other two
directions. Instead, the much higher resolution reachable in X-rays
experiments, is in principle adequate to perform the whole
3-dimensional Fourier transform of the spectrum. For concreteness
sake, we assume a Gaussian resolution with variance $\xi^2$, where
$\xi > R_a$ (the opposite case is not interesting as the resulting
peak shape is affected only by the resolution). We consider first the
behavior of the direct term $I_d$ in case of a RS solution. The peak
obtained from the $C^{RS}_{\text{d}}(r)$ function is essentially
independent of the resolution and has the shape of as a squared
Lorentzian of height $\propto R_a^3$ and half-width $\propto 1/R_a$.
The profile drastically change if we consider instead the correct RSB
solution.  Setting $\eta=1$ we can find an analytical expression for
the experimental peak:
\begin{eqnarray}
\label{last}
I^{\text{exp}}_d(K+Q+k)&\propto&  R_a^2 \int_r e^{-i k r}
e^{-\frac{r^2}{2 \xi^2}} \frac{1}{r} \\
&=&(2 \pi)^{3/2} \frac{R_a^2 \xi}{k} e^{-\frac{ k^2 \xi^2}{2}}
{\text erfi}(\frac{k \xi}{\sqrt{2}}) \,, \nonumber
\end{eqnarray} 
where ${\text erfi(x)}=-i {\text erf}(i x)$ is the imaginary part of
the error function. This expression gives a {\it non-divergent} peak,
shown in
\fig{ultima}, whose height is $4 \pi \overline{f}^2 K^2 {u_0}^2 R_a^2 \xi^2$
and its half-width $\sim 1.5034/\xi$.  We remark that the peak decays
with the characteristic law $q^{-2}$ only for $R^{-1}_a <q \ll
\xi^{-1}$, while a wide region around the maximum height is dominated
by the effects related to the finite resolution. In \fig{ultima} we
also report the function $I_d^{\text approx}(q) =I_d^{\text exp}(0)
(1- q^2 \xi^2 /3)$.  It is possible that, in a real experiment, the
signal-noise ratio is low enough to hide the $q^{-2}$ behavior. In
this specific case, to put in evidence the existence of a Bragg glass
regime one should vary the Fukuyama-Lee length $R_a$
\cite{klein_brglass_nature}.  In fact, if the peak shape is Lorentzian,
varying $R_a$ produces a change both in the width and the height of
the peak, whereas if the peak follows a power law, only its height is
changed due to a variation of $R_a$, while the width is fixed by the
resolution.

In absence of screening, the long range Coulomb interactions become
important: the system is more rigid and the upper critical dimension
is shifted from $d=4$ to $d=3$. As a consequence, the satellite peaks
become more divergent. In particular, we find that the symmetric term
$I_d$ goes asymptotically as $q^{-3}$ instead of $q^{-2}$.  Moreover,
the strong anisotropy between the longitudinal direction ($x$) and the
transversal ones ($y$) leads to an anisotropic scaling of the
correlation functions. We verified that in $d=3$ the dependence on $x$
and $y$ of the correlation functions (as clearly shown in
(\ref{chiani})) and of the characteristic length scale (\ref{larkin})
respects the relation $f(x,y=0)\sim f(x=0,y^2)$.  As a result, in this
regime, we expect more divergent and more anisotropic peak shapes 
in comparison to the ones observed in the 
short-range case.

\begin{figure}
\centerline{ \includegraphics[width=\figwidth]{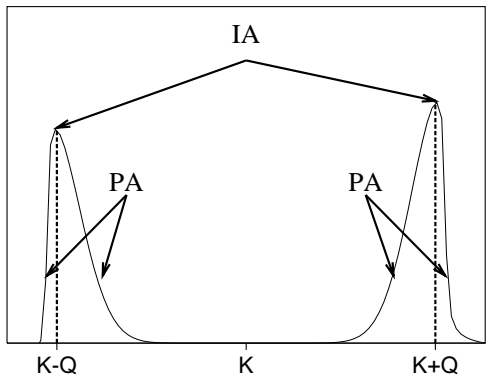} }
\caption{Sketch of the experimental findings
  \cite{rouzieres_structure_cdw}. The two satellites present an IA in
  agreement with our predictions, but also an evident PA.}
\label{sketch}
\end{figure}

On the experimental side few detailed diffraction spectra are
available at the moment. An example of prototype systems are doped
blue bronzes \cite{rouzieres_structure_cdw}, where the disorder is
introduced by partially replacing Mo$^{6+}$ for the non-isoelectric
V$^{5+}$.  In this kind of system, the interactions between the CDW
and the impurities are repulsive and $\Delta f$ is negative. The shape
of the observed spectrum is sketched in \fig{sketch}.  The sign of IA
is consistent with our prediction as well as with the predictions of
the strong pinning approach.  However, we also remark the presence of
PA of each single satellite. The mirror symmetry of the shape of the
two peaks suggests that the PA is originated in the correlation
function $C_d(r)$. As we have previously discussed, the correlation
function obtained from the FLR model with Gaussian disorder is real
and produces two peaks with profile {\it symmetry}.  One can wonder
what is the effect of non-Gaussian disorder. To this purpose, we
studied a $1$-dimensional model with a binomial distribution of
impurities. If we restrict to the forward scattering term in the
development (\ref{eq:rhoexp}), the problem is exactly soluble. We find
two Lorentzian peaks without PA, centered around a vector $K\pm
\tilde{Q}$ shifted with respect to $K\pm Q$ as a consequence of the
presence of odd moments in the disorder distribution.  It remains to
investigate the role possibly played by the amplitude fluctuations.
The short correlation lengths extracted from the experimental data
suggest that this particular effect is most likely to happen in the
strong pinning regime, whereas our calculations concern the weak
pinning limit.  The authors of
Ref.~(\onlinecite{rouzieres_structure_cdw}) indeed justify the PA with
the presence of Friedel oscillations and hence with the presence
strong fluctuations of the amplitude of the condensate, at least in
the neighborhood of the impurities. 

\section{Conclusions}

Summarizing, we determined the shape and the intensity of the
satellite peaks characterizing the spectrum of a pinned charge density
wave. We analyzed in detail the case of a weak and collective
disorder, when the Fukuyama-Lee-Rice model is justified. We considered
both the short-range elasticity as well as the long-range elasticity
generated by an unscreened Coulomb interaction. In both these cases,
we found divergent peaks displaying intensity asymmetry. The divergent
nature of the peaks is, as it was discussed, the clearest sign of a
Bragg glass phase. Moreover, the long-range elasticity, when present,
is responsible for a larger anisotropy and a stronger divergence. Let
us stress that the calculated sign of the intensity asymmetry is in
agreement with the experimental data. We discussed the role played by
the finite resolution of the experimental setup, calculating the
convolved shape of the peaks, where the divergence is cut. From these
observations, we illustrated possible methods to reveal experimentally
the presence of the Bragg glass phase.  Concerning the asymmetry of
the peak profile, we showed that, on general symmetry grounds, it is
not expected in the weak pinning regime. We conjectured that its
observation in a recent experiment \cite{rouzieres_structure_cdw} is
likely due to the strong pinning present in the measured system.
Finally, we observe that the profile asymmetry may hide the power law
behavior of the satellite peaks.  It would thus be highly desirable to
dispose of measures in less disordered systems where one can expect a
Bragg glass behavior, e.g. using isoelectric impurities.

\section{Acknowledgments}

We acknowledge the illuminating discussions with J.P.~Pouget,
S.~Ravy. This work was supported in part by the Swiss national fund
for scientific research.
\appendix

\section{\label{appendice} The triplet term}

In this appendix we discuss the behavior of the term $I_{\text{tripl}}$ of the
intensity development (\ref{eq:Intensities}).  This term was only conjectured to be
negligible \cite{ravy_x-ray_whiteline,cowley_x-ray_cdw}, before that 
\cite{rosso_cdw_short} a direct calculation of it was performed.

We start analyzing the symmetry properties of $I_{\text{tripl}}$ showing it
has the same symmetry of $I_{\text{a}}$.  Next we evaluate $I_{\text{tripl}}$
within the RS scheme.  Independently of the elasticity range, we find that the
difference between the two contributions is given by a simple $q$-independent
multiplicative factor.

Applying a Fourier Transform (\ref{FT}) we obtain
\begin{equation}
\label{tripletto1}
 I_{\text{tripl}}(K +\delta q) = -2iK \Delta f^2  a^{d} \int_{BZ}
 \overline{\langle \Sigma_{\delta q} \Sigma_{q_1} u_{-\delta q
 -q_1}\rangle} \,.
\end{equation}
This equation gives a non-zero contribution only if we consider higher
harmonic terms of the electron density (\ref{eq:rhoexp}): $\rho(x) \sim \rho_1
|\psi| \cos(Q(x+\phi(x))) +\rho_1 |\psi| \cos(2Q(x+\phi(x)))$.  The two
satellites take the form:
\begin{equation}
\label{tripletto2}
I_{\text{tripl}}(K + \delta q) = -2iK \Delta f^2  a^{d}
 \overline{\langle 
 \Sigma_{\delta q} [\Sigma_{- 2 \delta q} u_{\delta q}+\Sigma_{\delta
 q} u_{-2\delta q}]\rangle} \nonumber\\ 
\end{equation}
At this stage it is evident that $I_{\text{tripl}}$ has the same symmetry of
$I_{\text{a}}$. Performing the integration over the Gaussian disorder by means
of the standard replica techniques we obtain
\begin{equation}
\label{triplettoreplica}
 I_{\text{tripl}} = {\Delta f}^2 q  u_0  N_I a^d  D \int_r
 e^{-i \delta  q r} \left[ e^{-i  Q
 r} -c.c.\right] C_{\text{tripl}}(r)  \nonumber
\end{equation}
Using the usual decomposition $C_{\text{tripl}}(r) = e^{- \frac{ \tilde{B}}{2}}
  \chi_{\text{tripl}}(r)$ we write
\begin{eqnarray}
\label{Ctrip}
 \chi_{\text{tripl}}(r) = \frac{1}{n T^2}\sum_{a,b,c} & &\left[ e^{- T
     \int_{BZ} 
2[\tilde{G} -  G_{ac}]} \right.  \\
& & \left. e^{-T
\int_{BZ}[(\tilde{G} - G_{ab})+ (G_{bc} - G_{ab})]\cos qr }\right] .\nonumber
\end{eqnarray}
We introduce the  replica symmetric Ansatz. It is easy to check that for $n=0$ 
\begin{equation}
\frac{1}{n} \sum_{a b c} A_{a b c}  =A_{a a a} -\sum_{a \ne b} (A_{a a b}+A_{a
  b a}+A_{b a a}) + 2 \sum_{a \ne b \ne c}  A_{a b c} \nonumber
\end{equation}
Using this relation and (\ref{eq:RSGnd}) we can evaluate
$\chi_{\text{tripl}}(r)$.  In order to simplify the notation we recall that
${\phi^2}_T = 2 T \int_{BZ} G_c $ and $\chi(r)=[1 - exp(-T \int_{BZ} G_c cos
qr)]/T$.  From (\ref{Ctrip}) we obtain
\begin{equation}
 \chi_{tripl}(r) = \frac{\chi(r)}{T} \left[1- e^{- \frac{ \phi^2_T}{2} }
 \left( 2 \text{sinh}(T \int_{BZ} G_c cos qr)+1 \right) \right].  \nonumber
\end{equation}
Because we are interested in the long distance behavior we remark that
$\int_{BZ} G_c cos qr \to 0$, whenever $r\to \infty$. Developing up to the first
order we get
 \begin{eqnarray}
\label{chitriplfinal}
 \chi_{tripl}(r) &\sim& \chi(r) \frac{1- e^{- \frac{ \phi^2_T}{2} }
  }{T}   \nonumber \\
&\sim& \chi(r) \frac{\phi^2_T}{ 2 T} \,,
\end{eqnarray}
where the last step is valid at low temperature. In fact, in this regime
$C_{\text{tripl}}(r)$ and $C_{\text{a}}(r)$ have the same behavior in $r$.
If we take a spherical cut-off $\Lambda=2 \pi/a$, where $a$ is the lattice
space, the integral $\phi^2_T/(2 T)=(c \pi a)^{-1}$ is independent of the range
of the elasticity (for $a \to 0$).
We can compare the two terms $I_{\text{tripl}}$ and $I_{\text{a}}$, by writing the
ratio between the two intensities  
\begin{equation}
 \frac{I_{\text{tripl}}}{I_{\text{a}}} = - \frac{\Delta f}{
\pi \overline{f}} \sqrt{\frac{N_I a}{R_a}} \,.
\label{stima}
\end{equation}
For weak disorder $R_a \gg a$; it follows that $ I_{\text{a}} \gg
I_{\text{tripl}}$.

\begin{figure}
\centerline{ \includegraphics[width=\figwidth]{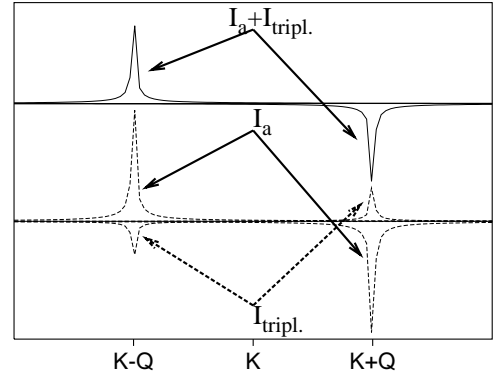} }
\caption{Intensities of the antisymmetric contributions to satellite
  peaks. We remark that
  $I_{\text{a}} \gg I_{\text{tripl}}$.}
\label{model}
\end{figure}

The evaluation of this term in a more accurate RSB approach is very complicate
because $\chi_{tripl}(r)$ involves the sum over three replicas. However, in
analogy with $\chi(r)$ we can argue that the RSB solution does not affect the
asymptotic power low behavior of $\chi_{tripl}(r)$.  In \fig{model} we
summarize our result taking the correct RSB behavior for $\tilde{B}$.

We conclude that the triplet term renormalizes the prefactor of $I_{\text{a}}$
without changing the power law behavior. In particular, when $\Delta f >0$,
$I_{\text{tripl}}$ enhances the asymmetry between the two satellites, while when
$\Delta f <0$, $I_{\text{tripl}}$ slightly decreases the antisymmetric
contribution.

\section{\label{appendix:sigma} Calculation of $[\sigma]$}

In this appendix we determine the variational function $[\sigma](v)$
defined in Eq.~(\ref{eq:sigmaapp}). We start form  the saddle point equation
(\ref{eq:replica1b}). From (\ref{eq:replica2}) we know that
\begin{eqnarray}
\label{eq:replicastart}
B(r=0,v) =  \int_{BZ}  \tilde{G}(q)-G(q,v) \,.
\end{eqnarray}
The integral in the momentum space is performed in the Brillouin zone.  To
simplify the analytical form of our integrals the ultraviolet cutoff,
$\Lambda$, is taken equal to infinity whenever it is possible ({\it i.e.}
whenever the integrals are ultraviolet convergent). Inserting (\ref{eq:RSBGnd})
in the previous equation and taking the derivative of (\ref{eq:replica1b}) leads
to the equation determining $[\sigma]$:
\begin{eqnarray}
\label{eq:sigma}
 \sigma(v)\int \frac{d^dq}{(2 \pi)^d} \frac{ T}{[G^{-1}_c +[\sigma]]^2}=1 \,.
\end{eqnarray}
Solving the integral for $2<d<4$ ($\Lambda \to \infty$) and deriving again, one gets
\begin{eqnarray}
\label{eq:sigmares}
 [\sigma](v)&=& \Sigma  (v/v_c)^{2/\theta} \;\;\; \mbox{for} \, v < v_c  \nonumber \\ 
  &=& \Sigma \;\;\;\;\;\;\; \mbox{for}\,  v > v_c \,,
\end{eqnarray}
where $\theta=2-d$. The values of the breakpoint $v_c$ and $\Sigma= c l^{-2}$
determine the crossover between a short distance regime ($x \ll l$), where
$[\sigma]$ is constant and the RS solution valid, and the asymptotic regime
($x > l$), where the physics is determined by the small $v$ behavior of
$[\sigma]$. Using Eqs.~(\ref{eq:sigma}) and (\ref{eq:replica1b}), after some manipulation
it is found \cite{giamarchi_vortex_long} for $d=3$
\begin{eqnarray}
\label{eq:cross}
l &=& \frac{1}{8 \pi} R_a e^{-\phi^2_T} \nonumber\\
v_c &=& \frac{T}{8 l c} \,. 
\end{eqnarray}
We observe that the crossover between the two regimes is $l
\sim R_a$ in agreement with  the dimensional result (\ref{eq:larkin}).  

For $d=4$, and more in general at the upper critical dimension, the integral
in (\ref{eq:sigma}) has a logarithmic ultraviolet divergence. As discussed in
Ref.~[\onlinecite{chitra_vortex_upper}], the behavior of $[\sigma]$ when $v$
is small is not described by a pure power law. Starting from (\ref{eq:sigma}) for
$d=4$ we get
\begin{eqnarray}
\label{eq:sigma4}
1 &=&\sigma(v)\int \frac{d^4q}{(2 \pi)^4} \frac{ T}{[c q^2 +[\sigma]]^2}
 \nonumber \\ &=& \frac{S_4 \sigma(v) Q^2}{c^2}
 \int^{\sqrt{c/[\sigma]}\Lambda}_0 
 \frac{q^3 d q}{[ q^2 +1]^2} \nonumber \\
 &\sim& \frac{S_4 \sigma(v) T }{2 c^2}\log(\frac{c \Lambda^2}{[\sigma]})  
\end{eqnarray}
Where $S_d=2^{1-d} \pi^{-d/2}/\Gamma(d/2)$ is the angular integration in $d$
dimension. Defining $A= 2 c^2/(S_4 T )$, we obtain after one more derivative
\begin{eqnarray}
\label{eq:sigma4final}
[\sigma]&=&\frac{Av}{\log^2\frac{[\sigma]}{c \Lambda^2}} \nonumber\\
        &\sim&\frac{Av}{\log^2\frac{A v }{c \Lambda^2}} \,.
\end{eqnarray}
This result is valid up to log-log corrections.
\begin{figure}
\centerline{\includegraphics[width=\figwidth]{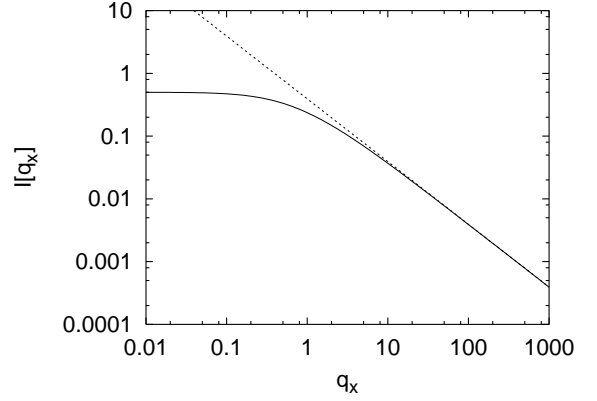} }
\caption{Continuum line: integral over q in (\ref{eq:sigmaoriginalb}) as a
  function of $q_x$. Dotted line: limiting behavior $I[q_x] \to \pi/8
  q_x$. Finally $I[0]=0.5$}
\label{fig:qint}
\end{figure}
From Eqs.~(\ref{eq:sigma4}) and (\ref{eq:replica1b}) we can estimate the crossover
length $l$. In the short distance regime, where $[\sigma]=c l^{-2}$, it turns out:
 \begin{eqnarray}
\label{eq:crossuc}
l &\sim& \Lambda^{-1} e^{ \frac{ 8 \pi^2 c^2}{D}} \,.  
\end{eqnarray}

At this stage we can discuss the case of the Coulombian long range elasticity.
Starting from Eq.~(\ref{eq:sigma}) and using Eq.~(\ref{eq:gclr}) for $G_c$, we write
 \begin{eqnarray}
\label{eq:sigmaoriginal}
1 &=& \sigma(v) \int^{\Lambda_x}_0 \frac{d q_x}{2
  \pi}\int^{\Lambda}_0\frac{d^2q}{(2 \pi)^2} \frac{T
 }{[\frac{c_1q^2_x+cq^4}{q^2}+[\sigma]]^2} \,. \nonumber 
\end{eqnarray}
To solve this equation we consider the physical case, where $\Lambda_x \sim Q$,
$\Lambda \sim 2 \pi /a$ and $a$ is the lattice space. In this limit
$\Lambda_x \ll \Lambda $ so we can assume $\Lambda\to \infty$ and solve the
integral
 \begin{eqnarray}
\label{eq:sigmaoriginalb}
1&=&\frac{ T\sigma(v) }{4 \pi^2 \sqrt{c^3 c_1}}
  \int^{\sqrt{c_1c}\Lambda_x/[\sigma]}_0 d q_x I[q_x] \nonumber \\
I[q_x] &=&\int^{\infty}_0 \frac{q^5 d q 
  }{\{q^2_x+q^4+1\}^2} \,. \nonumber
\end{eqnarray}
The behavior of  $I[q_x]$  is sketched in
\fig{fig:qint}. In conclusion, we obtain
\begin{eqnarray}
\label{eq:sigmalrfinal}
[\sigma]&=&\frac{A_{lr}v}{\log^2\frac{[\sigma]}{ \sqrt{c_1 c}\Lambda_x}}
        \nonumber\\ 
        &\sim&\frac{A_{lr}v}{\log^2\frac{A_{lr} v }{ \sqrt{c_1 c}
        \Lambda_x}} \,
\end{eqnarray}
where $A_{lr}= 16\pi\sqrt{c^3 c_1}/T$. This equation is equivalent to
the one found for an isotropic system at the upper critical dimension.


\begin{thebibliography}{34}
\bibitem[{eta()}]{eta_exponent_variation}
\bibinfo{note}{For a CDW the elastic object (the phase $\phi$) is a scalar. The
  exponent $\eta$ is universal and the best prediction is given by FRG
  calculation $\eta \sim 1.1$. However also in the case where the elastic
  object has many components (i.e. for the vortex lattice) the possible
  variation of this exponent is still less than a percent. [S. Bogner, T. Emig
  and T. Nattermann Phys. Rev. B {\bf 63} 174501 (2001)]}.
\end{thebibliography}

\end{document}